\documentclass[%
 reprint,
superscriptaddress,
 amsmath,amssymb,
aps,
]{revtex4-2}

\usepackage{graphicx}% Include figure files
\usepackage{dcolumn}% Align table columns on decimal point
\usepackage{bm}% bold math
\usepackage{textgreek}%pour mu droit
\usepackage{xcolor}
\begin{document}

\title{Shear thickening in presence of adhesive contact forces: the singularity of cornstarch}

\author{Anaïs Gauthier}
 \email{anais.gauthier@univ-rennes.fr, tel: +33223237300}
 \affiliation{%
MIE – Chemistry, Biology and Innovation (CBI) UMR 8231, ESPCI Paris, CNRS, PSL Research University, 10 rue Vauquelin, Paris, France}
\affiliation{%
Present address: Univ Rennes, CNRS, IPR (Institut de Physique de Rennes) - UMR 6251, F-35000 Rennes, France.}%

 \author{Guillaume Ovarlez}
\affiliation{%
 Univ. Bordeaux, CNRS, Solvay, LOF, UMR 5258, F-33608 Pessac, France}%

\author{Annie Colin}%
\affiliation{%
MIE – Chemistry, Biology and Innovation (CBI) UMR 8231, ESPCI Paris, CNRS, PSL Research University, 10 rue Vauquelin, Paris, France}%

\date{\today}

\begin{abstract}

\textit{Hypothesis:} A number of dense particle suspensions experience a dramatic increase in viscosity with the shear stress, up to a solid-like response. This shear-thickening process is understood as a transition under flow of the nature of the contacts— from lubricated to frictional — between initially repellent particles. Most systems are now assumed to fit in with this scenario, which is questionable.\\ 
\textit{Experiment:} Using an in-house pressure sensor array, we provide a spatio-temporal map of the normal stresses in the flows of two shear-thickening fluids: a stabilised calcium carbonate suspension, known to fit in with the standard scenario, and a cornstarch suspension, which spectacular thickening behavior remains poorly understood.\\
\textit{Findings:} We evidence in cornstarch a unique, stable heterogeneous structure, which moves in the velocity direction and does not appear in calcium carbonate. Its nature changes from a stress wave to a rolling solid jammed aggregate at high solid fraction and small gap width. The modeling of these heterogenities points to an adhesive force between cornstarch particles at high stress, also evidenced in microscopic measurements. Cornstarch being also attractive at low stress, it stands out of the classical shear-thickening frame, and might be part of a larger family of adhesive and attractive shear-thickening fluids.

\medskip
\noindent Keywords: shear thickening, adhesion, cornstarch, non-Brownian suspensions, normal stress measurement, phase separation.

\end{abstract}

\maketitle

\section{\label{intro} Introduction}

Many industrial processes, such as waste water treatment, oil drilling or mixing pastes require the manipulation of liquids containing a high solid fraction in particles \cite{Blanco:2019, Feys:2008}.
However, the flows of dense suspensions are particularly complex. One of their most striking features is shear thickening, where the viscosity can increase by more than one order of magnitude with the shear stress \cite{Barnes:1989,Brown:2014,Lootens:2003,Fall:2008,Guy:2015,Wagner:2009,Brown:2010}. For the densest suspensions, the material undergoes a liquid-solid transition: it is sometimes detrimental, causing pipe clogging or the breaking of blades in mixing systems \cite{Ovarlez:2020}. But it can also be conveniently used to engineer smart fluids \cite{gurgen2017shear, Ness:2018, Niu:2020, zarei2020application}, shock absorber for batteries \cite{liu2019shear}, damping devices \cite{zhang2008rheology} or body armors \cite{Mahfuz:2009, Majumdar:2014, xu2017stabbing, deGoede:2019}.

In this context, and in particular to propose new formulations~\cite{jiang2014study, gurgen2017shear, Hsu:2021}, a good understanding of the microscopic mechanisms at the origin of shear thickening is essential. In this field, the work of Wyart and Cates \cite{Wyart:2014, Mari:2014, Lin:2015, Royer:2016, Clavaud:2017} is central. Shear thickening is explained by the presence of a short-ranged \textit{repulsive} force between the particles. When the normal force exceeds the repulsive force \cite{Comtet:2017}, the particles switch from lubricated to frictional contacts, which leads to an increase of the viscosity. 

In the Wyart and Cates model, the suspension has a purely viscous behavior, with a viscosity $\eta$ depending on the ratio $\phi/\phi_j$ only (with $\phi$ the particle volume fraction and $\phi_j$ the jamming volume fraction) and diverging at $\phi/\phi_j$ = 1.  For the frictionless suspension, $\phi_j$ is the random close packing $\phi_{\mathrm{rcp}}$; for the frictional suspension, $\phi_j=\phi_m<\phi_{\mathrm{rcp}}$. In the model, $\phi_j$ depends on the fraction of particles in frictional contact; $\phi_j$ decreases from $\phi_{\mathrm{rcp}}$ to $\phi_m$ when the applied stress increases. With these ingredients, a critical solid volume fraction $\phi_{\mathrm{dst}}<\phi_m$ emerges and three regimes are found.
For $\phi<\phi_{\mathrm{dst}}$, the system shows continuous shear thickening (CST), characterized by a smooth transition from a low viscosity plateau (frictionless state with $\phi_j=\phi_{\mathrm{rcp}}$) to a high viscosity plateau (frictional state with $\phi_j=\phi_m$) when increasing the shear stress $\sigma$.
For $\phi_{\mathrm{dst}}<\phi<\phi_m$ the flow curve is sigmoidal, which points to hysteretic discontinuous shear thickening (DST) between the low viscosity state and the high viscosity state.
For $\phi_m<\phi<\phi_{\mathrm{rcp}}$ the material flows at low shear stress only and is jammed at high shear stress, a behavior coined shear jamming (SJ), as no flow is possible without dilatancy for the frictional material.  Above $\phi_{\mathrm{rcp}}$, the material cannot flow without dilatancy. This framework has been successful in capturing qualitatively the main features of shear thickening \cite{Morris:2018} for a wide range of suspensions, which now makes this description the standard explanation for shear thickening. Most systems are now assumed to fit in with this scenario, and focus is now made on the details of the transition.

Dense suspensions of cornstarch particles are the most emblematic shear-thickening systems, used both for scientific outreach \cite{Brown:2014} and in experiments designed to study the shear-thickening process \cite{Fall:2008, Hermes:2016, Ovarlez:2020, Boyer:2016, Saint-Michel:2018, Richards:2019, Darbois:2020, Niu:2020}. Most cornstarch suspensions show shear jamming \cite{Fall:2015,han2019stress}, with an apparent viscosity $\eta$ increasing by more than one order of magnitude at a critical shear rate. The rheological behavior of cornstarch has been fitted to the Wyart and Cates model in multiple occasions \cite{Hermes:2016, han2018shear, o2019liquid, Richards:2019, Darbois:2020}. A qualitative agreement was found close to the shear-thickening transition. The low shear-rate region, however, shows a yield stress \cite{Fall:2008}, which is not accounted for by the standard framework. Other comparisons with the Wyart-Cates theory have been performed with the help of phase diagrams in the ($\phi, \sigma$) parameter plane \cite{Fall:2015, han2019stress}, with a relative success. In cornstarch, the flow at high stress has been the object of a particular scrutiny in the past years. Unstable flows with large fluctuations are observed, with spatial and temporal heterogeneities of the velocity concentration and stress field in the gap of rheology cells \cite{Rathee:2022, Saint-Michel:2018, Ovarlez:2020, Richards:2019}. To this day, the exact nature and stability of these objects is still the object of research.

\smallskip

We propose here to revisit the question of the nature of the flow of cornstarch and of its origin. We focus in particular on the origin of the fluctuations. To do so, we have developed a new experiment that allows us to probe the normal stress -- a component of the flow that is rarely measured locally -- at the millimeter scale. We vary systematically the fraction of particles and gap size, which helps us understand the nature of the structure that is formed during the shear-thickening transition. We compare the local flow of cornstarch to that of another suspension (calcium carbonate), which microscopic and macroscopic properties follows the Wyart and Cates model. The comparison with this model fluid gives a novel insight on the shear-thickening process of cornstarch, which is shown to differ fundamentally from the often admitted frictionless to frictional mechanism.

\section{Materials and methods\label{Materials and methods}}

In this work, we compare the behavior of two fluids. The first one is the well-known dispersion of cornstarch particles (Merk). Here, and following Refs \cite{Fall:2008, Ovarlez:2020, Saint-Michel:2018}, the particles are placed in a density-matched liquid made of water and salt (Cesium Chloride, 55\% in weight). Cornstarch particles are polydisperse, with a mean diameter of 15 $\pm$ 7~\textmugreek m (see Supplementary Figure 1a). They are porous, so that a volume fraction cannot be determined without additional assumptions \cite{Han:2017}. Following Refs. \cite{Saint-Michel:2018,Ovarlez:2020}, we characterize the suspensions using the weight percentage in cornstarch $\phi_{\rm w} = \frac{m_p}{m_p + m_s}$ (with $m_p$ the mass of particles and $m_s$ the mass of solvent) instead of a volume fraction. In our experiments, $\phi_{\rm w}$ is varied between 34\% and 42\%. The second fluid is a calcium carbonate suspension (Eskal 500, KSL Staubtechnik GmbH) in a water-glycerol mixture. This system shares similarities with cornstarch: it is a polydisperse industrial powder, with particles having a rhomboidal-like shape and a mean size of 4 \textmugreek m (see picture in Supplementary Figure 1b). To our knowledge, stabilized calcium carbonate suspensions are one of the rare non-Brownian suspensions where two viscosity plateaus (at low and high stress) are actually observed \cite{Richards:2021}, together with silica suspensions \cite{Royer:2016, etcheverry2023capillary}. In particular, the flow curves can be quantitatively described by the Wyart and Cates model, over a wide range of shear rates and volume fraction \cite{Richards:2021}. Here, following Richards~\textit{et al.} \cite{Richards:2021}, the particles are dispersed in a 50-50\% water-glycerol mixture containing 0.05 w/w\% of polyacrylic acid. The poly-acrilic acid (PAA) removes the adhesive contacts between the grains: in presence of PAA, they interact as purely frictional hard particles with a short range repulsive force \cite{Richards:2021}, similarly to model systems. Contrary to cornstarch, calcium carbonate particles are not porous: we thus characterize the suspensions using the classical volume fraction $\phi$, defined as the volume of the solid particles divided by the total volume of the particles and solvent. It is varied between 35\% and 52\%.

\begin{figure}[!ht]
\centering
\includegraphics[width=0.99\columnwidth]{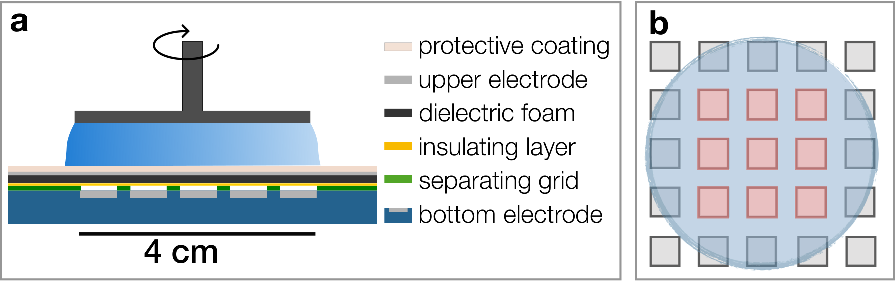}
\caption{\label{figure1} 
\textbf{a.} Normal stress measurement setup. A piezo-capacitive pressure sensor array is placed at the bottom of a parallel-plate rheometer. The diameter of the geometry is 4~cm and the gap thickness is typically 1~mm. The sensor is a sandwich of many layers (see legend). Its core is a soft house-made dielectric foam, which capacitance (measured in 25 points) depends on the local pressure. \textbf{b.} Top view of the experiment, where each square represent a different pressure sensor. The blue circle shows the geometry. Here, the pressure is measured in the nine positions shown in red.}
\end{figure}

We characterize the flow of the suspensions by combining classical rheology measurements with a mapping of the normal stresses at the millimeter scale, using an in-house sensor array. As shown in Figure~\ref{figure1}a, the sensor array is placed on the bottom plate of a conventional rheometer (DHR-2, TA instruments) in a parallel-plate geometry with a gap height $h$ varied between 0.5~mm and 3.0~mm. It consists of 25 regularly spaced piezo-capacitive sensors, built together in a 5 $\times$ 5 array of total width 4~cm. The details of the sensor array are shown in Figure~\ref{figure1}a. It is a sandwich of three main layers: \textit{i)} an electrode array (the bottom part), which sets the position and the size of the measurement points, \textit{ii)} a 25~\textmugreek m thick Mylar grid which separates the top and the bottom parts and \textit{iii)} a soft piezo-capacitive layer (made from a solid polymeric foam developed previously in our lab \cite{Pruvost:2019}), which is covered with a soft silver electrode. When a pressure is applied on the foam, it partially collapses and bends onto the grid, which impacts its local capacitance. By measuring the capacitance between the top and the bottom electrodes, we can thus map the vertical component of the stress in the fluid, with a spatial resolution of 4.5 $\times$ 4.5~mm$^2$ (the size of the electrodes), a precision of 2~Pa and a temporal resolution of typically 50~Hz \cite{Gauthier:2021}.

%%%%%%%%%%%%%%%%%%%%%%%%% FIGURE 1 %%%%%%%%%%%%%%%%%%%%%%%%
%%%%%%%%%%%%%%%%%%%%%%%%%%%%%%%%%%%%%%%%%%%%%%%%%%%%%%%%%%%%%%%

\smallskip

During an experiment, the pressure of the nine central sensors (in red in Figure \ref{figure1}b), which are fully covered by the geometry, is recorded. In a typical experiment, the shear stress $\sigma$ is imposed: it is increased by constant steps of duration 50 to 250~s, between 0.05 and 1000~Pa. The rheometer records the shear rate $\dot{\gamma}$ and the average vertical stress $P_{\rm m}$ through the built-in force sensor. Simultaneously, the sensor array on the bottom plate records the local normal stress $P$ as a function of time $t$. All these data are aggregated to obtain both a global and a local view of the shear-thickening process.

\section{Results and discussion
\label{Results and discussion}}
When discussing the flow curves of cornstarch, we use the term discontinuous shear-thickening to refer to situations where the flow curve is vertical or sigmoidal. We call continuous shear thickening the situation where the viscosity increases smoothly with the shear rate.

\subsection{Local flow in cornstarch suspensions
\label{Results}}

\begin{figure*}[!ht]
\centering
\includegraphics[width=0.9\textwidth]{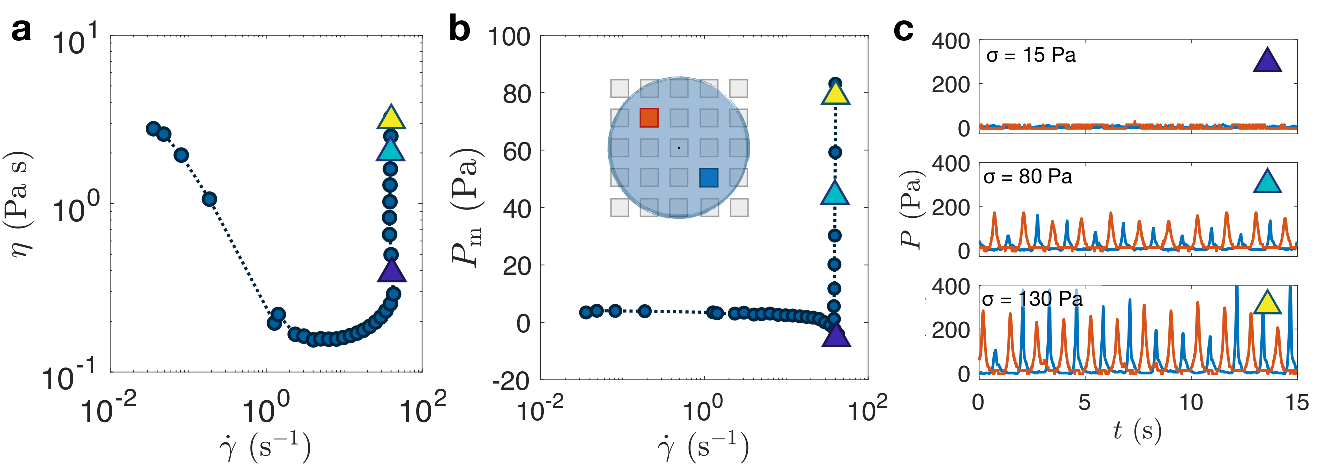}
\caption{\label{figure2} 
\textbf{a.} Flow curve (viscosity $\eta$ as a function of the shear rate $\dot{\gamma}$) of a cornstarch suspension with weight fraction $\phi_{\rm w} = 40$\%. \textbf{b.} Mean vertical pressure $P_{\rm m}$ as measured by the force sensor of the rheometer. \textbf{c.} Local normal stress as a function of time $t$, for increasing shear stresses $\sigma$. The inset of b. shows the position of the two measurement points in the sensor array: the red and blue sensors are placed symmetrically with respect to the center of the geometry. The position of the measurements in the flow curves are indicated with colored triangles.}
\end{figure*}

The flow curve of a cornstarch suspension with weight fraction $\phi_{\rm w}$ = 40 \% is presented in Figure \ref{figure2}a. Three different regions are distinguished: a shear-thinning region at low shear rates $\dot{\gamma} <$ 3~s$^{-1}$, a continuous shear-thickening region for 5~s$^{-1}$ $<\dot{\gamma} <$ 40~s$^{-1}$ and a discontinuous shear-thickening region for $\dot{\gamma} \simeq 40$~s$^{-1}$. At low shear rates, the strong shear thinning is consistent with a small yield stress of order 0.1~Pa, which matches previous measurements \cite{Fall:2008, Fall:2012}. As shown in Figure \ref{figure2}b, the mean pressure $P_{\rm m}$ measured by the force sensor of the rheometer is slightly negative and decreasing with $\dot{\gamma}$ in the first two regions of the flow curve. It switches to a positive value, and increases strongly with the shear stress in the discontinuous shear-thickening regime. This is characteristic of all the cornstarch suspensions that we have considered here, with with 35\% $< \phi_{\rm w} < 42$ \% (Supplementary Figure 2a). At these weight fractions, fluctuations of the macroscopic normal stress also appear in the discontinuous shear thickening region (see Supplementary Figure 2b and c) and their amplitude increases with the applied stress.

%%%%%%%%%%%%%%%%%%%%%%%%% FIGURE 2 %%%%%%%%%%%%%%%%%%%%%%%%
%%%%%%%%%%%%%%%%%%%%%%%%%%%%%%%%%%%%%%%%%%%%%%%%%%%%%%%%%%%%%%%

Using the sensor array, the flow is probed at the millimeter scale. For clarity, we present in Figure \ref{figure2}c the signal of two sensors (in red and blue) placed symmetrically under the geometry (see the schematic in the inset of Fig.~\ref{figure2}b) as a function of time. The signal measured by all nine sensors in the continuous shear thickening region (Supplementary Movie 1) and in the discontinuous shear thickening region (Supplementary Movie 2 and 3) is available in the Supplementary Materials. In the shear-thinning and in the continuous shear-thickening regions of the flow curve ($\sigma = 15$~Pa, top panel), the local normal stress matches the mean normal stress: the pressure measured by each of the sensors is constant and slightly negative. In the discontinuous shear thickening region ($\sigma = 80$~Pa and $\sigma = 130$~Pa), however, the two sensors present pressure peaks of short duration in phase opposition. As evidenced in Supplementary Movie 2, it corresponds to a unique pressure wave that rotates within the geometry in the velocity direction. The peak pressure is typically 4 to 5 times higher than the mean normal stress measured by the rheometer, and, between two events, the pressure is close to zero and slightly negative. When the shear stress is increased, for example from $\sigma$ = 80~Pa (middle panel) to $\sigma$ = 130~Pa (bottom panel), the amplitude of the stress signal increases, but we still observe a unique wave. This regular local normal stress wave is not evidenced by the force sensor of the rheometer, as shown in Supplementary Figure 2c and d. The mean normal stress only presents some noise around a constant value, significantly smaller than the local peak pressure. When increasing the applied stress, both the mean normal stress $P_m$ and the amplitude of the noise increase, while, at the small scale, the amplitude of the normal stress peak increases.

This single, stable and regular stress wave is observed in all cornstarch suspensions with a weight fraction $\phi_{\rm w} >$ 35\% and for gap widths varied between 0.5~mm and 2~mm. The regularity of the signal is lost in the situations of large gap width ($h \ge$ 2~mm) or small weight fraction (35\% $\leq \phi_{\rm w} <$ 37\%), as shown in Supplementary Figure 3. Finally, the signal completely disappears for $\phi_{\rm w} \leq $ 34\%, when the shear thickening is continuous (Supplementary Figure 4).

\begin{figure}[!ht]
\centering
\includegraphics[width=0.99\columnwidth]{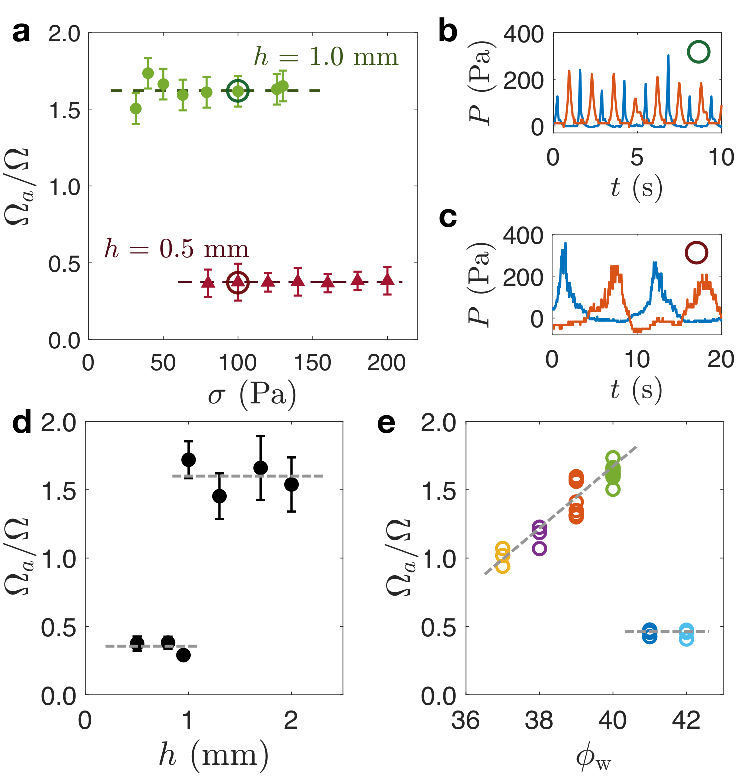}
\caption{\label{figure3} \textbf{a.} Non-dimensional velocity $\Omega_a/\Omega$ of the stress signal in a 40\% cornstarch suspension, as a function of the applied shear stress $\sigma$. The green circles correspond to a gap height of $h$ = 0.5~mm and the red triangles to $h$ = 1.0~mm. \textbf{b.} Normal stress signal $P$ measured by 2 sensors placed symmetrically as a function of time $t$ measured for $h$ = 1.0~mm, $\sigma$ = 100~Pa. \textbf{c.} Normal stress signal $P(t)$ for $h$ = 0.5~mm, $\sigma$ = 100~Pa. \textbf{d.} Non-dimensional velocity $\Omega_a/\Omega$ of the stress signal in a 40\% cornstarch suspension, as a function of $h$. \textbf{e.} Velocity of the stress signal as a function of the weight fraction $\phi_{\rm w}$ ($h$ = 1.0 mm).}
\end{figure}

The pressure wave being very regular, we now focus on its velocity. In Figure \ref{figure3}a we plot the angular velocity of the signal $\Omega_a$ (compared with the angular velocity $\Omega$ of the geometry) as a function of the applied shear stress $\sigma$. Interestingly, $\Omega_a/\Omega$ is independent of $\sigma$: it keeps a constant value $\Omega_a/\Omega = 1.6 \pm 0.1$ for $h$ = 1.0~mm (in red) and $\Omega_a/\Omega = 0.4 \pm 0.1$ for $h$ = 0.5~mm (in blue). What is remarkable here is that the two signals obtained for $h = 1.0$~mm and $h = 0.5$~mm, that could seem similar from a distance, surely correspond to very different physical situations. Indeed, the normal stress wave is faster than the geometry for $h$ = 1.0~mm ($\Omega_a > \Omega$): it is thus very unlikely that it originates from a solid object. However, for $h = 0.5$~mm, $\Omega_a \simeq 0.5\, \Omega$: the angular velocity of the signal is close to the mean velocity of the fluid within the geometry, which could indicate the presence of a rolling solid aggregate. The shape of the signal is also different in both situations: when $\Omega_a/\Omega > 1$, the signal is peaked as in Figure \ref{figure3}b. For $\Omega_a/\Omega < 1$ (Figure \ref{figure3}c, $h$ = 0.5~mm), the signal is more noisy and the width of the peak is larger.

%%%%%%%%%%%%%%%%%%%%%%%%% FIGURE 3 %%%%%%%%%%%%%%%%%%%%%%%%
%%%%%%%%%%%%%%%%%%%%%%%%%%%%%%%%%%%%%%%%%%%%%%%%%%%%%%%%%%%%%%%

To get a better picture of this phenomenon, we varied systematically the gap thickness $h$ (0.5~mm $<h<$ 2.0~mm, Figure \ref{figure3}d) and the weight fraction in particles $\phi_{\rm w}$ of the suspensions (37\% $\leq \phi_{\rm w} \leq$ 42\%, Figure \ref{figure3}e). In both cases, we evidence an abrupt transition between a slow pressure wave (with $\Omega_a/\Omega$ constant and close to 1/2 of the velocity of the geometry) and a fast wave (with a signal faster than the geometry: $\Omega_a/\Omega > 1$). The slow signal is measured when the suspension is confined ($h <$ 1~mm ) and dense ($\phi_{\rm w} \geq 41\%$), while the fast one is seen at lower weight fraction and large gap. The transition between the two regimes is sharp: for $\phi_{\rm w} = 40\%$, the velocity of the signal is multiplied by 3 when the gap width increases from 0.95~mm to 1.0~mm (Figure \ref{figure3}d); moreover, this velocity decreases by a factor $\simeq 3$ when switching from $\phi_{\rm w} = 40$\% to $\phi_{\rm w} = 41$\% for $h$ = 1~mm. It is also associated with a visible change in the shape of the signal recorded by the sensors, as evidenced in Supplementary Figure 5: similarly to what is shown in Figure \ref{figure3}c, the ``slow'' pressure signal is more noisy and of longer peak extent than the fast one.

\subsection{Comparison with calcium carbonate suspensions}

\begin{figure*}[!ht]
\centering
\includegraphics[width=0.99\textwidth]{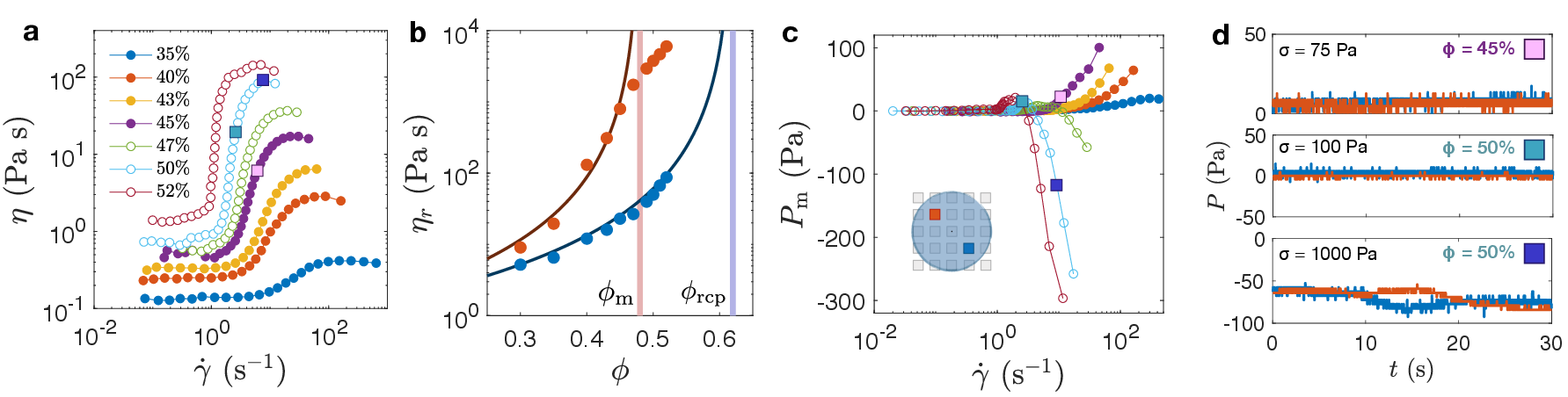}
\caption{\label{figure4} \textbf{a.} Flow curve of calcium carbonate suspensions, for volume fractions $\phi$ varied between 35\% and 52\%. The full circles correspond to the continuous shear-thickening regime, and the empty circles to the discontinuous shear-thickening regime. \textbf{b.} Relative viscosity $\eta_r = \eta/\eta_0$ in the low (in blue) or high (in red) plateau (with $\eta_0$ = 23 mPa\,s the viscosity of the suspending fluid). The blue and the red curves are fits of Krieger-Dougherty form \cite{Krieger:1959} $\eta_r = \left(1-\phi/\phi_{\rm j}\right)^{-\beta}$ with $\beta$ = 2.5. $\phi_j$~=~$\phi_{rcp}$~=~0.62 for the low viscosity plateau and $\phi_j$~=~$\phi_m$~=~0.48 for the high viscosity plateau. The vertical pink line materializes the divergence of the viscosity of the high viscosity branch at $\phi = \phi_m$; the vertical blue line shows the divergence of the viscosity of the low viscosity branch at $\phi = \phi_{\rm rcp}$. \textbf{c.} Mean vertical stress $P_m$, the color code is the same as in $\textbf{a}$. \textbf{d.} Stress signal measured by the sensor array, of 2 sensors placed symmetrically below the geometry. The position of the measurements in the flow curves are indicated with the colored squares.}
\end{figure*}

In order to know whether the presence of a stable normal stress wave is specific to cornstarch or not, we have considered another shear-thickening fluid: a calcium carbonate suspension in a water-glycerol mixture, stabilized by a very small quantity of poly-acrylic acid. As shown by \citet{Richards:2021} and Figure \ref{figure4}a, these suspensions behave as in the theory and the simulations of an ideal hard sphere system \cite{Wyart:2014, Mari:2014}. By contrast with cornstarch, calcium carbonate suspensions are not shear thinning: only two regions are distinguished, and the shear thickening happens between a low viscosity plateau and a high-viscosity plateau.

%%%%%%%%%%%%%%%%%%%%%%%%% FIGURE 4 %%%%%%%%%%%%%%%%%%%%%%%%
%%%%%%%%%%%%%%%%%%%%%%%%%%%%%%%%%%%%%%%%%%%%%%%%%%%%%%%%%%%%%%%

Here, and contrary to cornstarch, we can clearly define the  random close packing volume fraction $\phi_{\rm rcp}$ and the jamming volume fraction $\phi_{\rm m}$ from the flow curve. This is done in Figure \ref{figure4}b, where we report the relative viscosity $\eta_r = \eta/\eta_0$ in the low (in blue) and high (in red) viscosity plateaus, with $\eta_0 = 23$ mPa\,s the viscosity of the suspending fluid. The value of $\eta_r$ in the low-viscosity region is obtained by averaging all values of $\eta$ before shear thickening (typically for $\dot{\gamma}<$ 0.2 s~$^{-1}$), and on the last 4 points after shear thickening for the high viscosity region.

In Figure \ref{figure4}b, the viscosity $\eta_r(\phi)$ is fitted to a  Krieger-Dougherty model \cite{Krieger:1959} $\eta_r = \left(1-\phi/\phi_{\rm j}\right)^{-\beta}$, with $\beta$ = 2.5. From this, we can extract the low-shear jamming volume fraction ($\phi_j = \phi_{\rm rcp}$ = 0.62) and the high-shear jamming volume fraction ($\phi_{\rm j} = \phi_{\rm m}$ = 0.48), which are respectively materialized in Figure \ref{figure4}b by a vertical blue and pink line, and correspond to the divergence of the viscosity in the frictionless and frictional regions of the flow curve. From this, and similarly to \citet{richards2021turning}, we expect calcium carbonate to be in the continuous and discontinuous shear-thickening regime for $\phi < \phi_m$ (filled circles in Figure \ref{figure4}a and c) and in the shear jamming regime at higher volume fractions (open circles). This is further confirmed in Figure \ref{figure4}c, when plotting the mean normal stress. For $\phi <$ 47\%, $P_{\rm m}$ is positive and increases with $\dot{\gamma}$. For $\phi >$ 47\%, $P_{\rm m}$ is first positive and suddenly decreases (in the region of the apparent high viscosity plateau). Following \cite{Richards:2021} and \cite{Dhar:2019}, we interpret the drop away from the model in the high viscosity plateau in Figure \ref{figure4}b along with the sudden decrease of the normal stress as a fracturing of the jammed suspension for $\phi \geq 47$\% at high shear rate.

Using the normal stress sensors, we now focus on the variations of the local normal stress $P$ with time $t$ along the flow curves of calcium carbonate, measured for two sensors placed symmetrically below the geometry (see the insert of Fig. \ref{figure4}c). The top panel of Figure \ref{figure4}d shows a typical signal in the continuous shear-thickening regime ($\phi = 45$\%, pink square) and the middle panel shows it in the discontinuous shear-thickening regime ($\phi = 50$\%, light blue). In both cases, the pressure $P$ does not vary with time, a striking difference with what is observed in cornstarch suspensions. Here, $P$ is constant, positive and of the order of 10~Pa, in good agreement with the mean normal stress $P_{\rm m}$ (figure \ref{figure4}c). 
The bottom panel of Figure \ref{figure4}d shows the signal in the fracturing -- shear-jammed -- regime (dark blue square). In this region, the two sensors measure a different signal, which varies with a timescale of $\sim 10$~s. This is a bit larger than the period of rotation of the geometry, and could indicate slow rearrangements of the fracture within the jammed suspension. Here also, the signal is fundamentally different from anything that we have detected in cornstarch.

\subsection{Discussion}

The results obtained with the calcium carbonate suspension show that flow instabilities and fluctuations are not present in all shear-thickening fluids, despite being often reported \cite{Lootens:2003, Saint-Michel:2018, Richards:2019, Rathee:2017, Rathee:2022, Ovarlez:2020, Hermes:2016}. A stable flow in the shear-thickening regime, as seen for calcium carbonate is fully in line with theory and simulations, which predict that both homogeneous or inhomogeneous flows are possible \cite{Olmsted:2008, Hermes:2016, Chacko:2018}. In the case of inhomogeneous flows, what seems expected from the Wyart and Cates model is unstable bands moving in the \textit{vorticity} direction \cite{Chacko:2018}.

In the following discussion, we will focus on the local flow and the shear-thickening process in cornstarch. We will first consider the nature and origin of the stress wave. We will then see how this new result fits in with the current knowledge on cornstarch, by carrying out a critical analysis of the bibliography. We will finally see that cornstarch seems to belong to another family of shear-thickening fluids, which calls for an additional shear-thickening mechanism.

\subsubsection{On the nature of the stress wave in cornstarch}

In previous experiments, large temporal fluctuations of the shear rate or the viscosity have been reported \cite{Hermes:2016, Rathee:2022, Saint-Michel:2018, Ovarlez:2020, Richards:2019, Boersma:1991, Lootens:2003, Pan:2015, Bossis:2017}, associated with an heterogeneous flow \cite{Rathee:2017, Saint-Michel:2018, Ovarlez:2020, Rathee:2020}. Here, we evidence in cornstarch something of the same line: with the help of the pressure sensors, we clearly detect in the parallel-plane geometry a unique wave, moving in the velocity direction, which presence generates macroscopic fluctuations of the shear rate and normal stress. An original aspect of our work is to reveal that there are two categories of pressure signals, depending on the confinement and on the fraction of particles in the suspension. Interestingly, the existence of two different local flows reconciles two previous experiments performed separately on cornstarch, by Ovarlez \textit{et al} \cite{Ovarlez:2020} and Rathee~\textit{et al.} \cite{Rathee:2022}.

Indeed, using X-ray radiography in a 41\% cornstarch suspension, Ovarlez~\textit{et al.} see a single and regular wave moving at a constant speed $\Omega_a < \Omega$ in the velocity direction, associated with an object denser than the rest of the fluid. This signal is fully consistent with our experimental observations of the ``slow'' stress wave. Using the same suspension and the same gap width as in Ref. \cite{Ovarlez:2020} ($h$ = 0.5~mm; in a parallel-plate geometry instead of a Couette cell) we measure a normal stress signal with a velocity $\Omega_a$ = 0.40 $\Omega$, which is close to the value reported in Ref. \cite{Ovarlez:2020}, with $\Omega_a$ = 0.30 $\Omega$. By combining our results, we deduce that 
the signal that we detect at large weight fractions and small gap width is most likely associated with the motion of a high solid fraction object, associated with a local jamming of the suspension. This rolling floc would naturally move at a velocity $\Omega_a = 0.5\,\Omega$ or more slowly if it partially slides: this is exactly what we observe in our experiments (Figure \ref{figure3}) and what is shown in Ref. \cite{Ovarlez:2020}.

By contrast, at low weight fraction $\phi_{\rm w} \leq 40$\% and large gap height ($h>$ 1~mm), the signal that we detect is faster than the geometry, which is not consistent with the motion of an assembly of particles. We infer that, in this region, the signal is rather the signature of a density wave. This observation is compatible with the recent results of Rathee \textit{et al.} \cite{Rathee:2022}: by combining boundary stress microscopy and particle tracking, they evidenced (at a smaller scale than our experiment) the presence of stress waves associated with propagating fronts. Similarly to our experiment, the stress waves are moving in the velocity direction at a velocity $\Omega_a \simeq \Omega$ or slightly higher. However, a quantitative comparison is not possible: in Ref \cite{Rathee:2022}, the authors add glycerol in the solution, which has been shown to modify the interparticle forces \cite{Galvez:2017}.

\smallskip

\begin{figure}[!ht]
\centering
\includegraphics[width=0.9\columnwidth]{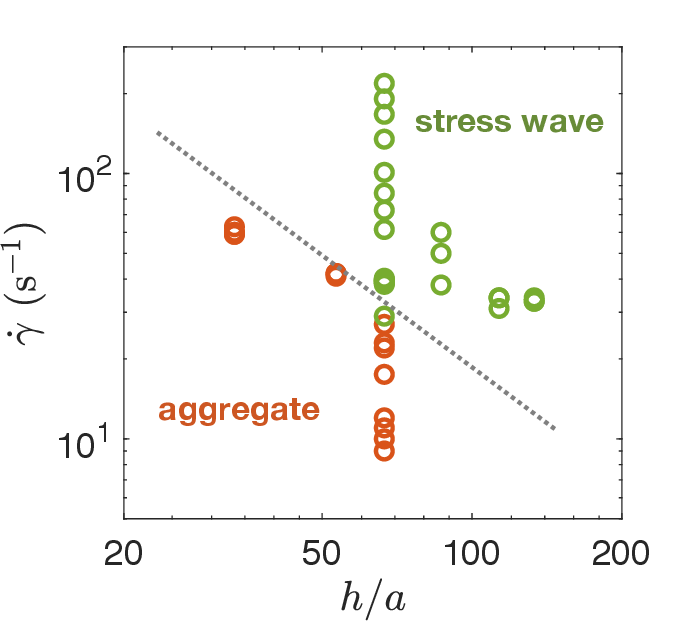}
\caption{\label{figure5} Phase diagram showing the two different normal stress signals (in red, the signal associated with a possible aggregate and in green, with a normal stress wave) as a function of the shear rate $\dot{\gamma}$ and the gap width $h$ (compared with the particle size $a$ = 15 \textmugreek m. The dotted line corresponds to the prediction of Ref. \cite{Varga:2019} on the stability of log-rolling flocs, without any adjustable parameter.}
\end{figure}

\color{black} How can we now explain the change in the nature of the signal with the experimental conditions? \\
Log-rolling flocs, which could correspond to what we see at high fraction of particles and small gap, are often seen in dilute suspensions of weakly attractive or adhesive particles sheared between parallel planes. The conditions of formation and of stability of these objects has been studied recently by Varga~\textit{et al.} \cite{Varga:2019}. They have shown that log-rolling flocs can only appear when the viscous drag force acting on them (which tends to break them) is smaller than the total interaction force between the particles within the floc. Thus, a stable phase separation in a system of attractive or adhesive particles is only possible at low enough shear rate and gap height \cite{Varga:2019}. This is a feature that we also see in cornstarch, as evidenced in Figure \ref{figure5}. In this phase diagram, we show the region in which the possible aggregate is seen (in red) as a function of the shear rate and dimensionless gap height $h/a$, with $a$ the mean particle size. The green dots show the cases where a stress wave is detected. There is a clear transition between the two regimes, the aggregate being only present at low enough $\dot{\gamma}$ and $h/a$. 

 %%%%%%%%%%%%%%%%%%%%%%%%% FIGURE 6 %%%%%%%%%%%%%%%%%%%%%%%%
%%%%%%%%%%%%%%%%%%%%%%%%%%%%%%%%%%%%%%%%%%%%%%%%%%%%%%%%%%%%%%%

These results suggest that cornstarch particles are attractive and/or adhesive at high stress. A way to verify this hypothesis, and to estimate the typical interaction force between the particles is to determine the stability region of the flocs, in our experiment. To do so, we follow the approach of Varga~\textit{et al.} \cite{Varga:2019}. At the boundary between the stress waves and the aggregates, we expect the objects to have the same size as the gap, and roll at half the speed of the top plate. In a shear flow, an aggregate of diameter $h$ is subjected of a torque $M = 6\pi\eta\dot{\gamma}h^3$ that destabilizes it. It is stable only if the yield torque $\Gamma$ associated with its internal structure is capable of resisting it. $\Gamma$ is directly proportional to the microscopic force $F$ associated with the attractive or adhesive bonds between the particles. It writes $\Gamma = n_l F a$, with $n_l \sim \left(h/a\right)^{d_f -1}$ the number of links in a plane of the fractal structure and $a$ the size of a particle. Thus, the condition of stability of the particles $M < \Gamma$ writes $6\pi\eta\dot{\gamma}h^3<(h/a)^{d_f-1}aF$, which can be expressed as a function of the Mason number: $\text{Mn} < 2.5\left(\frac{h}{a}\right)^{d_f - 4}$, with $\text{Mn} = \frac{6\pi\eta\dot{\gamma}a^2}{F}$. $d_f$ is taken to be 2.6 in the \citet{Varga:2019} experiments, so the transition criterion is $\text{Mn} < 2.5\left(\frac{h}{a}\right)^{-1.4}$.

In Figure \ref{figure5}, we compare our experimental results to the model (shown with a dotted gray line) with $a$ = 15~\textmugreek m. In the flow curves of cornstarch, the shear-thickening regime (where the pressure waves are present) appears at a critical shear rate $\dot{\gamma}$, which value decreases sharply with the weight fraction $\phi_{\rm w}$ (Supplementary Figure 2a). For this reason, the transition between a stress wave and an aggregate when increasing the weight fraction at a fixed gap height $h$ (Figure 3e) directly translates as a $\dot{\gamma}$ transition. On figure 5, a transition at high shear rates correspond to low weight fractions, while a transition at low shear rates is associated with a high fraction of particles. In this ($\dot{\gamma}$, $h/a$) phase diagram, the model fits nicely our experimental measurements, by taking an interparticle force at contact $F$ = 20 nN, a value consistent with an earlier measurement of the force between two particles using atomic force microscopy \cite{Galvez:2017}. This result has two important consequences: first, it shows that the results of \citet{Varga:2019} can apply to  dense suspensions, where the aggregates cannot be visually observed, but can be detected by other means. In addition, and more importantly, it shows the existence of an attractive or an adhesive force between the cornstarch particles at high shear. As discussed later, other papers lift the uncertainty on the nature of the force, and clearly point towards an adhesive force (present after contact between the particles) instead of an attractive force that could be measured before contact. To our knowledge, this is the first time that an interparticle force can be indirectly measured \textit{in the flow}. Indeed, such forces are usually measured using atomic force microscopy (AFM) \cite{Galvez:2017, James:2018, Hsu:2021}, by approaching a particle glued to the tip of the AFM to another particle fixed to a substrate. This method is very accurate but it is limited: it cannot fully mimic what happens under shear, since the dynamics of the approach and retraction between two particles in the flow are often too fast to be reproduced with this tool.

\subsubsection{Microscopic and macroscopic behavior of cornstarch}

The presence of adhesion at high stress that we evidence here is particularly meaningful when it is combined to what is already known about cornstarch.

\smallskip

Cornstarch particles are made of two biopolymers (amylose and amylopectin) that are partially soluble in water \cite{Buleon:1998,Green:1975}. When dispersed in water, the granules are thus surrounded by a soft polysaccharide layer, which is evidenced when probing the microscopic interaction between particles. This has been done with quartz-tuning fork atomic force microscopy \cite{Comtet:2017} and classical atomic force microscopy (AFM) \cite{Galvez:2017}. Before contact, the interaction force between the starch grains is very small \cite{Comtet:2017,Galvez:2017}. The force profiles do not evidence any attraction, and repulsive forces, if any, are smaller than 2~nN. In the Wyart and Cates model \cite{Wyart:2014}, a 2~nN  repulsion force would imply a critical stress for the shear-thickening of order 1~Pa: this is more than one order of magnitude lower than the 20~Pa stress measured experimentally. This points to a first discrepancy with the Wyart and Cates scenario: if cornstarch was following the classical model, the expected repulsive force would be clearly observable, as \textit{e.g.} for PVC particles \cite{Comtet:2017}. Furthermore, while there is no measurable \textit{attractive} force (that would be visible when approaching the particles), a strong \textit{adhesive} force between the particles is measured. When the particles are put in contact with an indentation force of the order of 100 nN, an hysteresis is observed at the retraction, which indicates the presence of an adhesive force of the order of $10 - 20$~nN (in water) \cite{Galvez:2017}. This force extends over typically 100 nm after contact. When separating the particles, a series of "pulling events" have been evidenced, corresponding to small and sudden jumps in the force profile \cite{Galvez:2017}. Such events are seen when high density dangling polymers (in a poor solvent) disentangle, and corresponds to the breaking of bonds during the retraction \cite{Yu:2015}. These results indicate that cornstarch particles interact through adhesive forces at high stress. They are most likely slightly swollen,  \cite{Belitz:2008} and surrounded by a collapsed polymer brush, a microscopic configuration that is known to generate adhesion \cite{Hsu:2021}.

The absence of a significant repulsive force implies that cornstarch particles are already in solid contact at low stress, in opposition to what is expected from the Wyart and Cates model. This is supported by three other observations. First, cornstarch suspension is a yield stress fluid: it has been evidenced in particular by \citet{Fall:2008}, which measured the velocity profiles of a cornstarch suspension in a Couette Cell. From the shear-localization of the flow, they measured a yield stress $\tau_y$ of the order of 0.1 Pa. This points to a small attractive force between the particles at low shear, of the order of $\tau_y\pi(a/2)^2 \simeq$ 0.03 nN, far below the resolution on the force measurements of \cite{Galvez:2017,Comtet:2017}. Second, the avalanche angle measurements performed by \citet{Clavaud:2017}, which have been successful in showing that potato starch and silica particles are in a frictionless state at low shear stress, failed to show the same feature for cornstarch. The results are not shown in Ref. \cite{Clavaud:2017} but can be found in the PhD dissertation \cite{clavaud2018rheoepaississement}, where aggregate formation and high value of the avalanche angle are reported (see e.g. Fig.~2.10 in \cite{clavaud2018rheoepaississement}), suggesting that the low shear stress state is not frictionless. Third, it should be mentioned that pressure-imposed rheology, which has been successful in demonstrating the frictionless/frictional transition in potato starch and silica beads when increasing the particle pressure \cite{etcheverry2023capillary}, failed to demonstrate the same behavior for cornstarch. The results are not shown in Ref. \cite{etcheverry2023capillary} but can be found in the PhD dissertation \cite{etcheverry2022capillarytron} where high values of the macroscopic friction coefficient are reported for dense slow flows at low particle pressure (see eg. Fig.~4.3 in \cite{etcheverry2022capillarytron}), suggesting again that the low shear stress state is not frictionless.

It should finally be noted that most papers that report an agreement between the rheological behavior of cornstarch and the Wyart and Cates model only fit the flow curves in a very limited range of shear rates, close to the shear thickening. The low shear-rate region is often disregarded, as well as the high shear rate limit. Both regions actually show a shear stress plateau as reported by \citet{Fall:2015}, who show the flow curves for a wide range of shear rates and volume fractions.

From all these previous results combined, we point out the presence of attractive forces at low stress in cornstarch and the absence of a frictionless/frictional transition with increasing stress. Our experimental results add another important information to this: cornstarch particles are also in contact at high stress. In this region of the flow curve, the adhesive force is of the order of 20 nN, which is 1000 times higher than the attractive force at the origin of the yield stress.

Cornstarch thus appears as a singular shear-thickening fluid, in which particles are in contact all along the flow curve, due to attractive or adhesive forces. For this reason, it cannot fit in with the Wyart-Cates framework, which assumes repulsive short-range forces between the particles to explain shear thickening. Macroscopically, the flow curves also differ from the model. Indeed, Most papers that report an agreement between the rheological behavior of cornstarch and the Wyart and Cates model only fit the flow curves in a very limited range of shear rates, close to the shear thickening. The low shear-rate region is often disregarded, as well as the high shear rate limit. Both regions actually show a shear stress plateau as reported by \citet{Fall:2015}, who show the flow curves for a wide range of shear rates and volume fractions.

\subsubsection{Shear thickening in presence of adhesive forces}

From the previous discussion, we see that another interpretation has to be found to explain the flow curve of cornstarch. We suggest here that cornstarch particles are very weakly attractive in suspensions in CsCl brine (to account for the yield stress), and strongly adhesive once the normal stresses pushes them into contact. More precisely, the initial contact between particles induces a small yield stress. When subjected to stress, the fragile bonds break and the suspension turns to a liquid, exhibiting a strongly shear thinning behavior. Beyond a critical stress, the particle pressure pushes the particles in contact and involves an important indentation. The polysaccharide chains surrounding the starch granules become entangled. An adhesive force coming from hydrogen bonds is set up. In this regime, there is a competition between the time for the polysaccharide to disentangle and the contact time ($\propto1/\dot\gamma$) between the particles. When the first is larger than the latter, the system thickens.

\medskip
\section{Conclusion}

% A summary of your findings.

To better understand the mechanism of shear thickening of cornstarch solutions, we study the structure of the suspensions under flow, with a focus on the high shear region. Using in-house sensors, we probe at the millimeter-scale, in a new way, the normal stresses that develop in two shear-thickening fluids: calcium carbonate and cornstarch. We show that the model suspensions of calcium carbonate particles (which are known to interact repulsively at low stress) remains homogeneous at all shear stresses, with a macroscopic flow curve that matches the Wyart and Cates model. In cornstarch, however, the sensors evidence a regular normal stress signal in the shear thickening regime. The nature of this signal changes with the constraining conditions: at low weight fractions $\phi_{\rm w} \leq$ 40\% and large gap widths ($h \ge$ 1~mm) the signal is consistent with a stress wave moving 1 to 1.5 times faster than the geometry. At high weight fractions $\phi_{\rm w} \ge$ 41\% and small gap ($h<1$~mm), the signal arises from the passage of a solid aggregate rolling between the plates. This transition is explained by taking into account the presence of \textit{adhesive} forces between the particles \textit{at high stress}. The presence these forces, combined with the existence of a yield stress (showing that the particles are also in contact at low stress) clearly indicates that cornstarch particles are in contact all along the flow curve.

% A synopsis of your new concepts and innovations.

To the best of our knowledge, this is the first time that the behavior of cornstarch suspensions at high shear stress is quantitatively described as a function of the gap height, shear rate and fraction of particles in the parameter plane ($h/a, \dot\gamma$). By mapping the boundary of the phase separation region as a function of the gap width and shear rate, we are able to use the presence of the solid aggregate as an original way to estimate the force of adhesion between the particles at high shear stress. The adhesion force measured at the transition is in perfect agreement with AFM measurements \cite{Galvez:2017}.

% A comparison with findings by other workers [give references]. 

% & vision of future work

Our work evidences the need for a new mechanism in the current research on shear thickening. Indeed, a known effect of attractive forces is to reduce or hide the shear thickening \cite{Brown:2010,pednekar2017simulation}. To our knowledge, a shear-thickening transition in presence of adhesive forces is not mentioned in the abundant literature on non-Brownian suspensions \cite{Brown:2014, Fall:2008, Guy:2015, Lin:2015, Mari:2014, Morris:2018, Royer:2016, Clavaud:2017, Comtet:2017, Hermes:2016, Niu:2020, Rathee:2022} or hydrocolloids \cite{Wagner:2009, Jamali:2015, Maranzano:2002}. \citet{Singh:2020} show that cohesive suspensions can present a shear thickening flow curve, but the simulations also include a repulsive force that we do not see in cornstarch. Indeed, the current understanding of the mechanism is that a repulsive force must somehow exist and be overcome during the flow to lead to an increase of the viscosity \cite{Wyart:2014, Mari:2014, Morris:2018}. While this vision works for a number of shear-thickening fluids \cite{Guy:2015, Lin:2015, Royer:2016, Richards:2019, Comtet:2017, Singh:2020}, we show here that it does not explain all occurrences of shear thickening. Strikingly, fluids such as the most famous one, namely cornstarch, in which particles are always in contact, can also exhibit typical shear-thickening flow curves.

We hope that this new insight will help to get a more complete picture of the shear-thickening process. This also brings to light a central issue: if cornstarch is far from being a canonical system, it should be used with care especially when its macroscopic flow is compared to the standard models and to simulations. In this line of work, our results also open new roads in suspensions formulation \cite{zarei2020application, gurgen2017shear}, as they highlight the possibility of engineering a new kind of adhesive shear thickening suspensions, which viscosity can be controlled through surface chemistry \cite{Hsu:2021}. To finish with, our work highlights the considerable importance of the microscopic interaction between the particles. In the past years, new measurements methods and original processes have been developed to control and measure the inter-particle force \cite{Comtet:2017, Galvez:2017, James:2018, Hsu:2021}. In the future, a better knowledge of both the microscopic forces and the dynamics between particles seem necessary to understand the details of the flow at all scales.

\begin{acknowledgments}
 We thank Vikram Rathee, Jeffrey Urbach, Daniel Blair and Joia Miller for fruitful discussions, and Mickaël Pruvost for his help in the fabrication of the sensors and the acquisition system.
\end{acknowledgments}

\medskip

\textbf{Credit authorship contribution statement}\\
\textbf{\textit{Anaïs Gauthier:}} Investigation, Formal analysis, Writing - original draft, review \& editing. \textbf{\textit{Guillaume Ovarlez:}} Conceptualization, Writing - review \& editing. \textbf{\textit{Annie Colin:}} Conceptualization, Writing - review \& editing, Resources, Project administration

\bibliography{biblioMaizena}% Produces the bibliography via BibTeX.

\end{document}